\documentclass[12pt]{elsart}
\usepackage{graphics}
\usepackage{epsfig}
\usepackage{setspace}
\usepackage{longtable}

\begin{document}
\begin{frontmatter}
\title{Hysteresis from dynamically pinned sliding states}
\author{Andrea Vanossi$^a$\corauthref{cor}},
\ead{vanossi.andrea@unimore.it} \corauth[cor]{Corresponding
Author.}
\author{Giuseppe E. Santoro$^{b,c}$},
\author{Nicola Manini$^{d}$},
\author{Marco Cesaratto$^{d}$},
and \author{Erio Tosatti$^{b,c}$}
\address{$^a$CNR-INFM National Research Center S3 and
Department of Physics, \\University of Modena and Reggio Emilia,
Via Campi 213/A, 41100 Modena, Italy}
\address{$^b$International School for Advanced Studies (SISSA)
and INFM Democritos National Simulation Center, Via Beirut 2,
I-34014 Trieste, Italy}
\address{$^c$International Centre for
Theoretical Physics (ICTP), P.O.Box 586, I-34014 Trieste, Italy}
\address{$^d$Department of Physics, University of Milan, Via Celoria 16, 20133 Milan, Italy}

\maketitle
\begin{abstract}
We report a surprising hysteretic behavior in the {\it dynamics}
of a simple one-dimensional nonlinear model inspired by the
tribological problem of two sliding surfaces with a thin solid
lubricant layer in between. In particular, we consider the
frictional dynamics of a harmonic chain confined between two rigid
incommensurate substrates which slide with a fixed relative
velocity. This system was previously found, by explicit solution of
the equations of motion, to possess plateaus in parameter space
exhibiting a remarkable quantization of
the chain center-of-mass velocity ({\it dynamic pinning}) solely
determined by the interface incommensurability. Starting now from
this quantized sliding state, in the underdamped regime of motion
and in analogy to what ordinarily happens for static friction, the
dynamics exhibits a large {\it hysteresis} under the action of an
additional external driving force $F_{\rm ext}$. A critical threshold
value $F_c$ of the adiabatically applied force $F_{\rm ext}$ is required in
order to alter the robust dynamics of the plateau attractor. 
When the applied force is decreased and removed, the system can jump to
intermediate sliding regimes (a sort of ``dynamic'' stick-slip motion) 
and eventually returns to the quantized sliding
state at a much lower value of $F_{\rm ext}$. 
On the contrary no hysteretic behavior is observed as a function of the external
driving velocity.
\end{abstract}

\end{frontmatter}

\section{Introduction}

Nonlinear systems driven far from equilibrium exhibit a very rich
variety of complex spatial and temporal behaviors \cite{Kapitaniak99}. 
In particular, in the emerging field of
nanoscale science and technology, understanding the nonequilibrium
dynamics of systems with many degrees of freedom which are pinned
in some external potential, as is commonly the case in solid state
physics, is becoming more and more often an issue. 
Friction belongs to this category too, because the microscopic
asperities of the mating surfaces may interlock. 
It has been frequently shown \cite{Vanossi_review} that simple
phenomenological models of friction give good qualitative
agreement with experimental results on nanoscale tribology or with
more complex simulation data of sliding phenomena. In this kind of
simplified approaches, studies are typically restricted to
describing microscopic dynamics in one (1D) or two (2D) spatial
dimensions. The substrates defining the moving interface are
modelled in a simplified, although often effective, way as purely rigid
surfaces or as one- or two-dimensional arrays of particles
interacting through simple (e.g., harmonic) potentials. Despite
this crude level of description, this approach has
frequently revealed the ability of modelling the main features of
the complex microscopic dynamics, ranging from regular to chaotic
motion.

One of the pervasive concepts of modern tribology -- with a wide
area of relevant practical applications as well as fundamental
theoretical issues -- is the idea of free sliding connected with
{\it incommensurability}. When two crystalline workpieces with
lattices that are incommensurate (or commensurate but not
perfectly aligned) are brought into contact, the minimal force
required to achieve sliding, i.e. the static friction, should
vanish, provided the two substrates are stiff enough. In such a
geometrical configuration, the lattice mismatch can prevent
asperity interlocking and collective stick-slip motion of the
interface atoms, with a consequent negligibly small frictional force. 
Experimental observation of this kind of {\it superlubric}
and anisotropic regime of motion has recently been reported
\cite{Frenken,Salmeron}. The remarkable conclusion of frictionless
sliding can be drawn, in particular, in the context of the
Frenkel-Kontorova (FK) model (see \cite{Braunbook} and references
therein). Since however the physical contact between two solids is
generally mediated by so-called ``third bodies'', the role of
incommensurability has been recently extended \cite{Braun} in the
framework of a driven 1D confined model inspired by the
tribological problem of two sliding interfaces with a thin solid
lubricant layer in between. The moving interface is thus
characterized by {\it three} inherent length scales: the periods
of the bottom and top substrates, and the period of the embedded
solid lubricant structure. In particular, in the presence of a
uniform external driving, the interplay between these
incommensurate length scales can give rise to intriguing dynamical
phase locking phenomena and surprising velocity quantization
effects \cite{Vanossi,Santoro}.

Extending a previous study of this confined tribological model
\cite{Vanossi}, here we focus on the remarkable hysteretic
behavior that this system exhibits starting now from this
quantized sliding state. We find a strictly analogy to what
ordinarily happens for {\it static} friction in the {\it
underdamped} regime of motion \cite{Braun97}. The lubricant
center-of-mass (CM) velocity turns out to be robustly locked to
the quantized plateau value (dynamic pinning) which is only
abandoned above a critical force. As long as inertia effects are
not negligible compared to dissipative forces, the adiabatic
variation (increase and decrease) of the external applied force
shows a large hysteresis loop in the $V_{\rm cm}$ -- $F_{\rm ext}$
characteristics. Some differences between this dynamic locking and
the usual static pinning are also briefly discussed.

\section{Confined model and numerical method}

Like in Ref. \cite{Vanossi}, we consider a simplified
one-dimensional generalized FK model consisting of two rigid
sinusoidal substrates, of spatial periodicity $a_{+}$ and $a_{-}$,
and a chain of harmonically interacting particles, of equilibrium
length $a_0$, mimicking the sandwiched lubricant layer, as
schematically shown in the inset of Fig.~\ref{Fig1}. The two
substrates move at a constant relative velocity $V_{\rm ext}=V_- -
V_+$. (In particular we set in full generality $V_+ = 0$ and $V_-
= V_{\rm ext}$). In order to probe the robustness of
quantized dynamics of $V_{\rm cm}$, an
additional constant force $F_{\rm ext}$ is applied adiabatically to
all chain particles. The equation of motion of the $i$-th
lubricant particle becomes:
\begin{eqnarray} \label{eqmotion:eqn}
m\ddot{x}_i &\,=\,& -\frac{1}{2} \left[ F_+ \sin{\frac{2\pi}{a_+}
x_i} + F_- \sin{\frac{2\pi}{a_-} (x_i-V_{\rm ext}t)}\right]
\nonumber \\
&& \hspace{1mm} + \, K (x_{i+1}+x_{i-1}-2x_i) 
-\, \gamma (2\dot{x}_i - V_{\rm ext})+ F_{\rm ext} \;,
\end{eqnarray}
where $m$ is its mass. $F_{\pm}$ are the amplitudes of the forces
due to the sinusoidal corrugation of the substrates. Presently we
set $F_-/F_+=1$ as the least biased choice. $K$ is the chain
spring constant defining the harmonic nearest-neighbor
interparticle interaction. The penultimate damping term in
Eq.~(\ref{eqmotion:eqn}) originates from two frictional contributions of the form 
$-\gamma\,(\dot{x}_i - V_+) - \gamma\,(\dot{x}_i - V_-)$, where $\gamma$ is a viscous friction
coefficient accounting phenomenologically for degrees of freedom
inherent in the real physical system (such as substrate phonons,
electronic excitations, etc.) which are not explicitly included in
the model. The infinite chain size is managed -- in the general
incommensurate case -- by means of periodic boundary conditions
(PBC) and finite-size scaling (see, for example, Refs.~\cite{Vanossi,Santoro}).
We finally take $a_+=1$, $m=1$, and $F_+=1$ as basic dimensionless units.

The detailed behavior of the driven system in
Eq.~(\ref{eqmotion:eqn}) depends crucially on the relative
(in)commensurability of the substrates and the chain. The
relevant length ratios are defined by $r_{\pm}=a_{\pm}/a_0$; we
assume, without loss of generality, $r_->r_+$, focusing mostly
on the case $r_+ > 1$. In particular, in order to make a
comparison with previous tribological studies \cite{Vanossi,Santoro}, 
we shall restrict our present analysis 
to the incommensurate golden-mean case $\phi\equiv(\sqrt{5}+1)/2\approx 1.6180$ 
with ratios $(r_+,r_-)=(\phi,\phi^2)$. Since the qualitative features of
velocity quantization phenomena were proved \cite{Vanossi} to survive for much 
more general values of $r_+$ and $r_-$, this specific choice of incommensurability 
should not be considered too restrictive.

The equations of motion (\ref{eqmotion:eqn}) are integrated using
a standard fourth-order Runge-Kutta algorithm. The system is
initialized with the chain particles placed at rest at uniform
separation $a_0$. After relaxing the starting configuration and
selecting a reference frame in which the bottom substrate is at
rest ($V_+=0$), the top substrate starts sliding at the imposed
constant velocity $V_-=V_{\rm ext}$. For a very wide range of model
parameters the system reaches, after an initial transient, the
quantized ``dynamical stationary'' state with
$V_{\rm cm}/V_{\rm ext} = V_{\rm plateau}/V_{\rm ext} = 1
- r_+^{-1}$, that depend solely on the chosen incommensurability
ratio $r_+$. In order to investigate the possibility for the
system to exhibit hysteresis starting now from this quantized
sliding state, at which the confined layer is robustly pinned, the
additional constant force $F_{\rm ext}$ acting on all chain particles
is now varied upward and downward adiabatically.

\section{Results and discussion}

\begin{figure}[p]
\centerline{
\epsfig{file=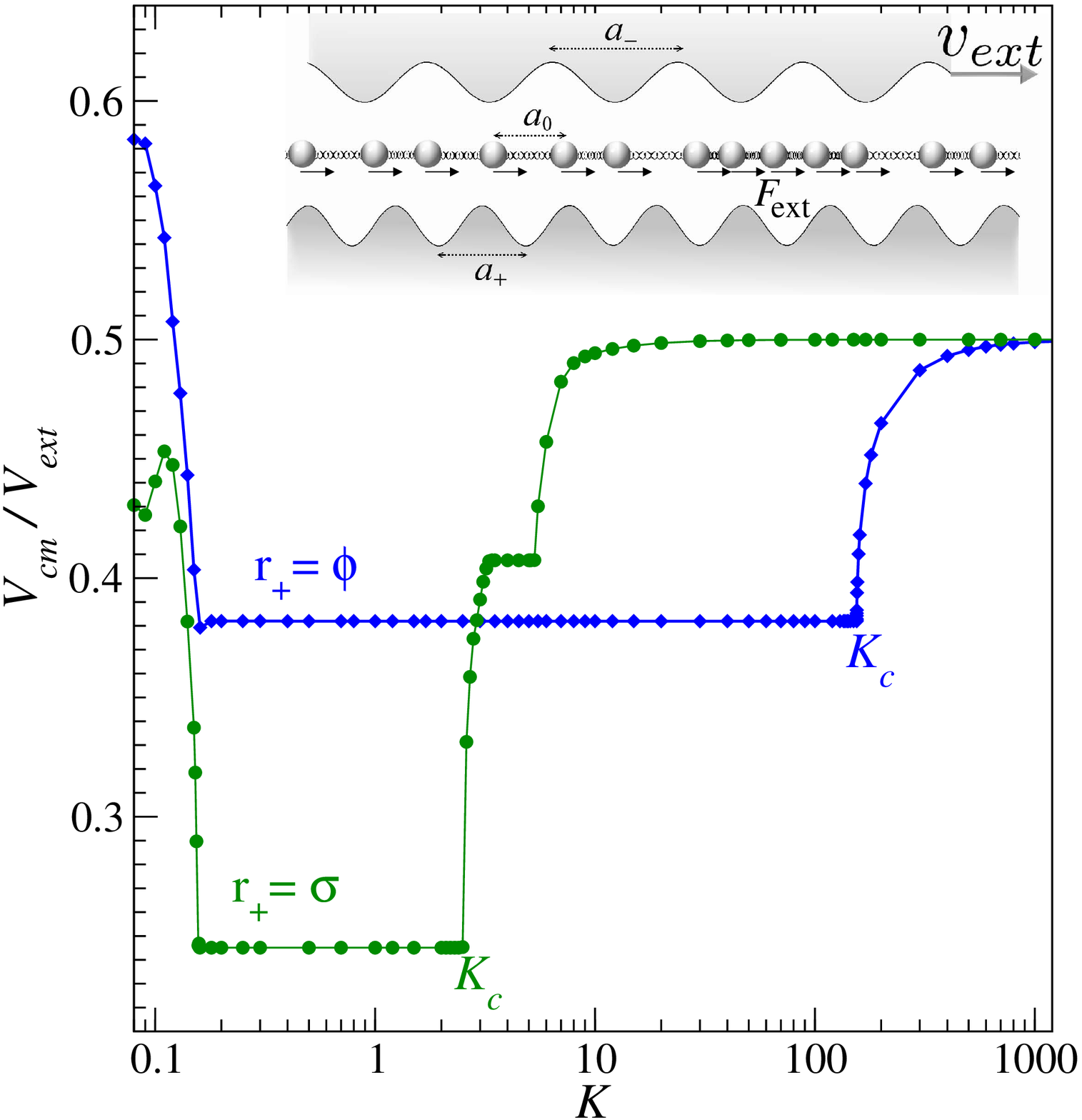,width=15cm,angle=0} } 
\caption{\label{Fig1}
(Color online) Normalized velocity of the center of mass, $V_{\rm cm}/V_{\rm ext}$, as a
function of the chain stiffness $K$, for the golden mean
$(r_+,r_-)=(\phi,\phi^2)$ and
spiral mean  $(r_+,r_-)=(\sigma,\sigma^2)$ incommensurability. 
Here $\gamma=0.1$ and $V_{\rm ext}=0.1$. Note the logarithmic scale in the abscissa. 
A sketch of the driven 3-length scale confined model is shown in the inset.}
\end{figure}

After stationarity has been reached, Fig.~\ref{Fig1} shows the
striking behavior of the normalized time-averaged CM velocity of
the sandwiched lubricant chain, $V_{\rm cm}/V_{\rm ext}$, as a function of
the stiffness $K$, for the two incommensurate case of golden mean
(GM) and spiral mean (SM)
\footnote{The golden mean $\phi\equiv(\sqrt{5}+1)/2$ 
is the solution of the quadratic equation $\phi^2-\phi-1=0$; 
the spiral mean, $\sigma\approx 1.3247$ satisfying the equation $\sigma^3-\sigma-1=0$, 
belongs to the class of cubic irrationals.}.

As discussed in previous works \cite{Vanossi,Santoro}, the crucial
feature is the presence of perfectly flat $V_{\rm cm}/V_{\rm ext}$
plateaus, whose precise value $(1-r_+^{-1})$ is independent not
only of $K$, but also of $\gamma$, $V_{\rm ext}$, and even of
$F_-/F_+$. Their occurrence was ascribed to the intrinsic
topological nature of this quantized dynamics. The phenomenon is
explained by one confining substrate rigidly dragging the
topological solitons (kinks) that the embedded chain forms with
the other substrate.

Let us now turn to consider specifically the GM case only.
Fixing a value of the chain stiffness $K$ lying approximatively in
the middle of the plateau of Fig.~\ref{Fig1} and considering a
sufficiently small value of the damping coefficient $\gamma$
(underdamped regime), we start investigating the hysteresis by
applying an external force $F_{\rm ext}$ to all the chain particles
through an adiabatic increase and decrease process.

\begin{figure}
\centerline{
\epsfig{file=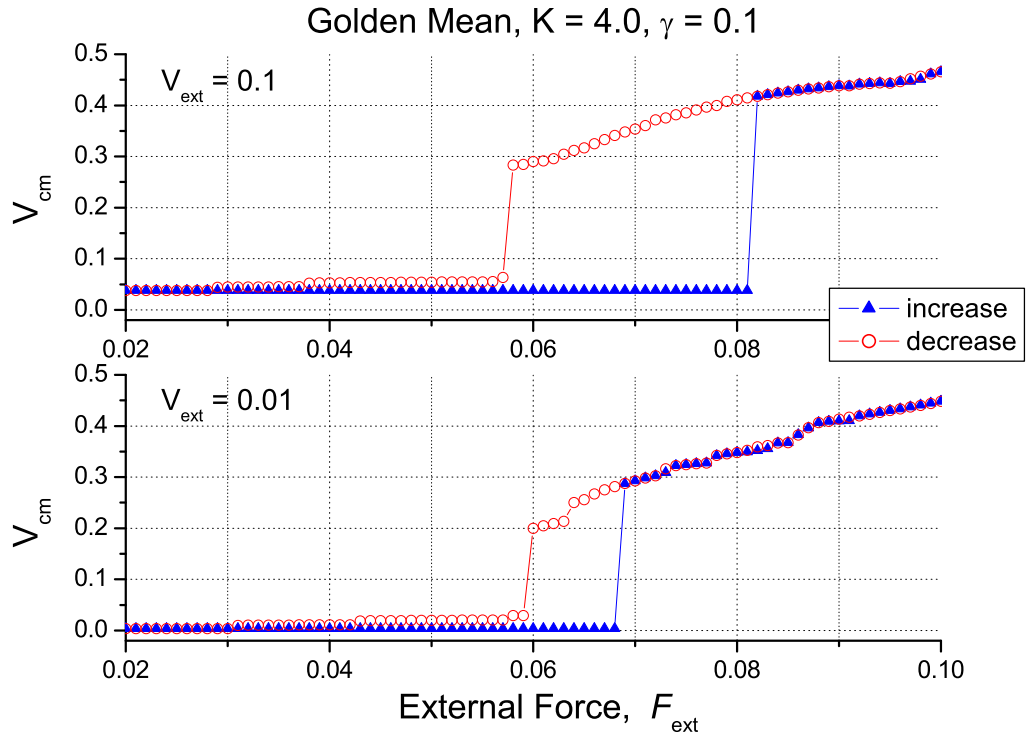,width=15cm,angle=0} } 
\caption{\label{Fig2}
(Color online) Hysteresis in the $V_{\rm cm}-F_{\rm ext}$ characteristics for the GM case
and a relatively soft ($K=4$) confined chain. The behavior is
shown for high ($V_{\rm ext}=0.1$, upper panel) and low
($V_{\rm ext}=0.01$, lower panel) applied driving velocities.
Adiabatic increase and decrease of $F_{\rm ext}$ is denoted by
triangles and circles, respectively. A characteristic multi-step
feature appears when decreasing adiabatically the external force.}
\end{figure}

The results are displayed in Fig.~\ref{Fig2} for two different
external driving velocities. A clear hysteretic loop emerges, with
qualitative similar features for high (upper panel) and low (lower panel) 
$V_{\rm ext}$. Surprisingly, the cycle is broader for larger velocities. 
We will return to this point later on.

\begin{figure}
\centerline{
\epsfig{file=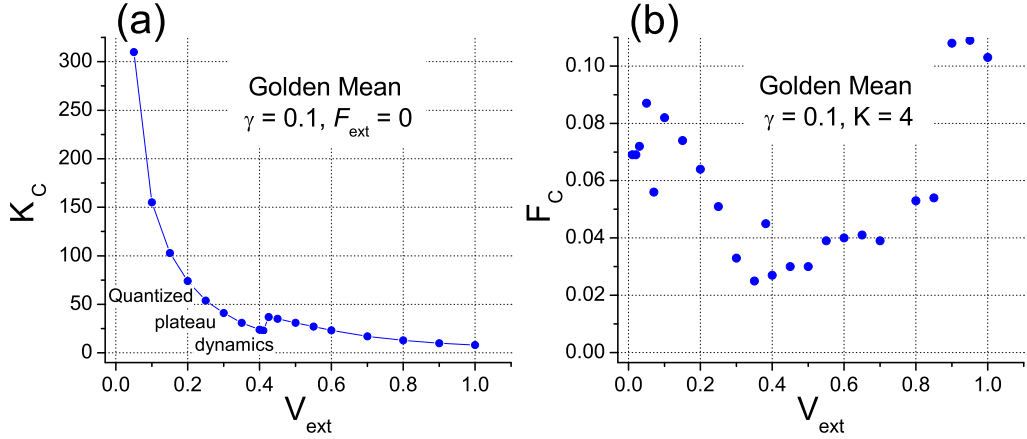,width=15cm,angle=0} } 
\caption{\label{Fig3}
(Color online)
({\bf a}) $(V_{\rm ext};K)$ phase diagram, with no additional force
applied, for the GM incommensurability. At a fixed value of the
external driving velocity, $K_c$ marks the point above which the
quantized sliding regime of the chain breaks down. 
({\bf b}) Irregular dependence of the dynamical depinning force $F_c$ upon
the driving velocity $V_{\rm ext}$ for chain stiffness $K=4$.}
\end{figure}

The finding of exact plateaus implies a kind of ``dynamical incompressibility'', 
namely, identically null response to
perturbations or fluctuations trying to deflect the CM velocity
away from its quantized value. What is now the effect of the
additional force $F_{\rm ext}$? We find that as long as $F_{\rm ext}$
remains below a critical threshold $F_c$, it does perturb the
single-particle motions but has no effect whatsoever on $V_{\rm cm}$,
which remains exactly pinned to the quantized value, as could
indeed be expected of an incompressible state. This picture is
analogous to the pinning-depinning transition in static friction,
where a minimum force (the static friction) is required in order
to start the motion. Thus the sudden change of $V_{\rm cm}$ taking
place at $F_{\rm ext}=F_c$ can be termed a ``dynamical depinning''.
The value of $F_c$ is a nontrivial function of the parameters, and
vanishes linearly when $K$ approaches from below the upper border $K_c$ of the
plateau. 
The depinning transition line $F_c$, ending at $K=K_c$, appears as a ``first-order''
line, with a jump $\Delta V$ in the average $V_{\rm cm}$ and a clear hysteretic
behavior as $F$ crosses $F_c$. 
As expected, we find that $\Delta V$ decreases to $0$ as $K$ increases towards $K_c$ (not shown).
Thus $K=K_c$ represents a sort of non-equilibrium critical point, where
the sliding chain enters or leaves a dynamical orbit.
The precise value of $K_c$ depends on parameters such as $V_{\rm ext}$ and $\gamma$;
however, its properties do not.
Fig.~\ref{Fig3}(a) displays the general decreasing behavior of $K_c$, 
and thus the corresponding diminishing extension of the plateau, as a function of the
external driving $V_{\rm ext}$ (no applied force $F_{\rm ext}$). 
As shown in Fig.~\ref{Fig3}(b), and contrary to what could be intuitively
expected, no straightforward relation seems to exist between the
plateau extension in $K$ and its robustness against the external
perturbing force $F_{\rm ext}$. At a fixed chain stiffness, quite
large values of $F_c$ are found even for very high sliding velocities, 
where the plateau extension has already been reduced significantly.

\begin{figure}
\centerline{
\epsfig{file=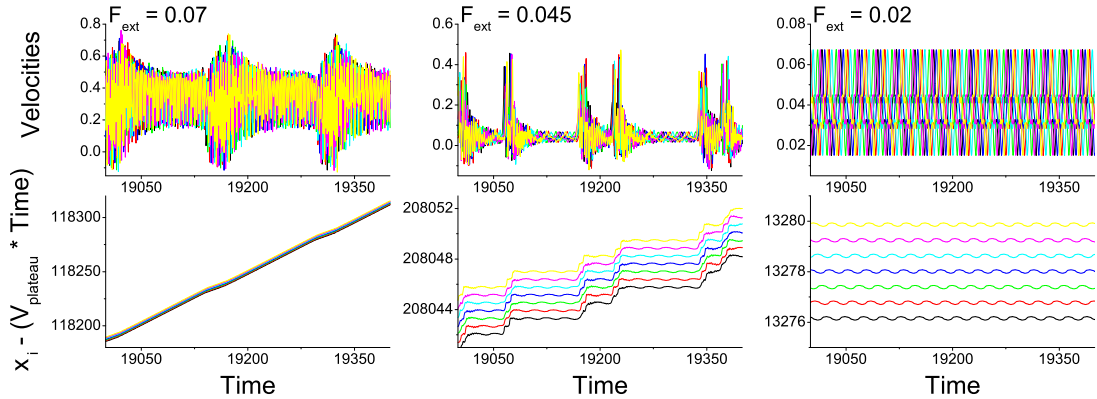,width=16cm,angle=0} } 
\caption{\label{Fig4}
(Color online)
Time evolution of the velocities $v_i$ (upper panels) and of the
corresponding rescaled coordinates $x_i - (V_{\rm plateau}\cdot t)$
(lower panels) of seven contiguous chain particles. The plots
refer to three different dynamical regimes observed in the
adiabatic decrease process of the external force: free sliding at
$F_{\rm ext}=0.07$ (left), dynamic stick-slip at $F_{\rm ext}=0.045$
(middle), and quantized sliding state at $F_{\rm ext}=0.02$ (right).
Note the different scale in the ordinate of the $v_i$-plots. 
Here $\gamma=0.1$, $K=4$, and $V_{\rm ext}=0.1$.
}
\end{figure}

Depending on model parameters such as chain stiffness and external
driving velocity, the dynamical depinning off the quantized
sliding state takes place through different kinds of mechanisms,
ranging from a series of intermittencies with a well-defined
temporal periodicity to more chaotic and irregular jumps \cite{Vanossi}. As
displayed in Fig.~\ref{Fig4}, when the applied force is decreased
and removed, the system may jump to intermediate sliding regimes
and eventually returns to the quantized sliding state at a much
lower value of $F_{\rm ext}$. As for the forward dynamical depinning
transition, the details of these backward steps in the $V_{\rm cm}$ --
$F_{\rm ext}$ characteristics strongly depend on the parameters of
the model. At high values of $F_{\rm ext}$($=0.07$, left panels) the
confined layer moves almost freely with a large sliding velocity
depending on the damping coefficient $\gamma$. For
intermediate force ($F_{\rm ext}=0.045$, middle panels), simulations
reveal the intriguing occurrence of a sliding regime closely
reminiscent of a dynamic stick-slip motion. This intermittent
dynamics is seen with particular clarity by plotting the particle
trajectories in the reference frame which slides at the quantized
velocity value of the plateau. A further reduction of
$F_{\rm ext}$($=0.02$, right panels) brings the system back to the
time-periodic dynamics of the quantized sliding state.

\begin{figure}
\centerline{
\epsfig{file=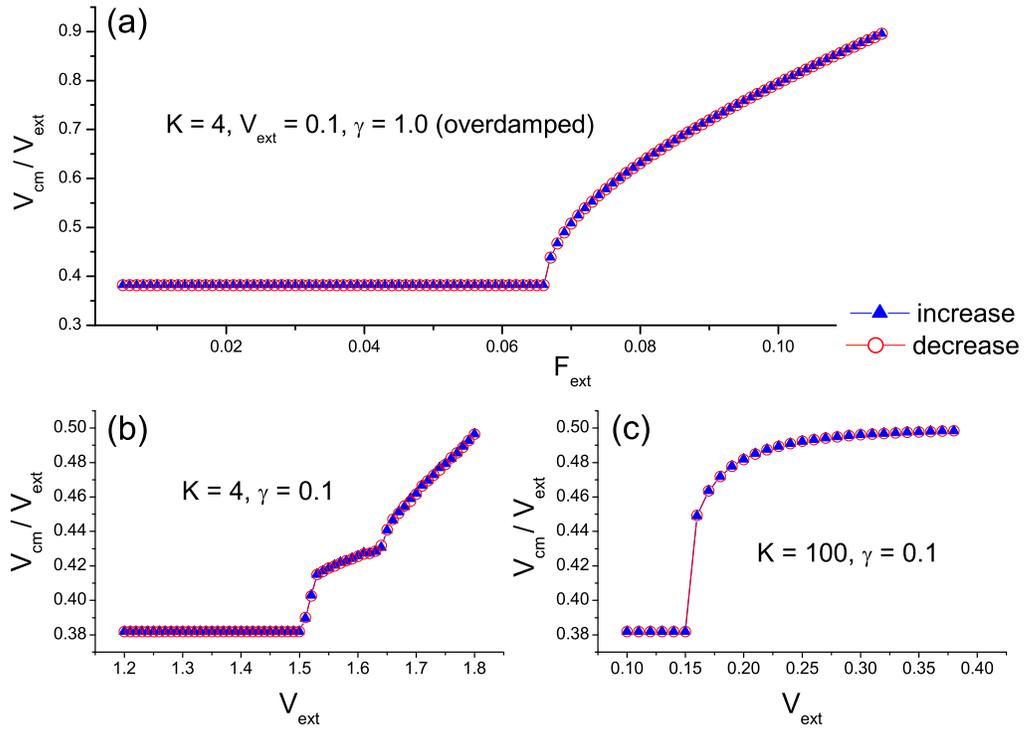,width=15cm,angle=0} } 
\caption{\label{Fig5}
(Color online)
No hysteretic behavior of $V_{\rm cm}/V_{\rm ext}$ is observed
in leaving the $V_{\rm cm}$ quantized plateau as a function of:
(a) the external applied force in the overdamped regime;
(b)-(c) the driving velocity, for different values of $K$.}
\end{figure}

The above picture of dynamical depinning as a first order
transition is valid for weak dissipation. For strong dissipation, when
the viscous damping coefficient $\gamma$ is much larger than the
characteristic vibrational frequencies of the system (overdamped
motion) the dynamical depinning is likely to be of second order:
The forward and backward trajectories become indistinguishable, 
and hysteresis disappears, as shown in Fig.~\ref{Fig5}(a). 
In this strongly dissipative regime, we found
instead of the hysteretic jumps a nonlinear mobility region of
$V_{\rm cm}$ versus $F_{\rm ext}$, but without any bistability phenomenon.

Finally, independently of the value of the chain stiffness, no
hysteresis has been found in the underdamped regime by varying the
external driving velocity with time with a gentle enough rate of
increase and decrease. Panels (b) and (c) of Fig.~\ref{Fig5} show,
for two different $K$ inside the quantized plateau region, the
nonlinear, but not bistable, behavior of $V_{\rm cm}/V_{\rm ext}$ as a
function of adiabatic variation of $V_{\rm ext}$.

\section{Conclusions}
%
We have shown that starting from the quantized $V_{\rm cm}$ sliding
state, previously found for a simple tribological model of a
confined layer, the layer sliding dynamics exhibits a large
hysteresis under the action of an additional external driving
force $F_{\rm ext}$ trying to change $V_{\rm cm}$ away from its quantized
value. In analogy to depinning in ordinary static friction, the
hysteretic dynamical behavior depends strongly on whether the
system degrees of freedom have sufficient inertia (underdamped
regime) or if, on the contrary, the inertia is negligible (overdamped regime).

The robustness of quantized dynamics is proved by the existence of
a finite critical threshold $F_c$ needed to move the chain CM
velocity away from the plateau value (dynamical depinning). When
the applied force is decreased and removed, the system may jump to
intermediate sliding regimes (a sort of dynamic stick-slip motion)
and eventually returns to the quantized sliding state at a much
lower value of $F_{\rm ext}$.

There are however nontrivial differences from static friction. The
first is that the dynamical pinning hysteresis cycle may be larger
in situations where the pinning itself could be intuitively
considered more fragile, e.g., for larger external velocity.
Another feature (presently under investigation, not discussed above)
is that the {\em sudden} application of an external force can leave $V_{\rm cm}$
locked to the quantized value, even if the applied force is
larger than the dynamic depinning threshold $F_c$, obtained instead
through the adiabatic procedure sketched above.
Once again, this is different from static depinning, 
usually requiring smaller force (than $F_s$) if applied suddenly.

A final open question concerns the effect of a finite temperature
on the dynamical hysteresis. Preliminary results obtained through a Langevin dynamics
indicate that, so long as the thermal energy is much smaller than an effective dynamical 
barrier (gap) preserving the incompressible plateau state, the velocity quantization
is still observed. We expect, therefore, that the qualitative dynamical hysteretic 
behavior should not change too much. These aspects are currently under investigation.

\section*{Acknowledgments}
This research was partially supported by PRRIITT (Regione Emilia
Romagna), Net-Lab ``Surfaces \& Coatings for Advanced Mechanics
and Nanomechanics'' (SUP\&RMAN) and by MIUR Cofin 2004023199, FIRB
RBAU017S8R, and RBAU01LX5H.


\end{document}